\documentclass[11pt]{article}
\usepackage{geometry}   
\usepackage{dsfont}
\usepackage{amsthm}
\usepackage{amsmath}
\usepackage{amssymb}
\usepackage{esint}
\usepackage{graphicx}
\usepackage{mathrsfs}             
\usepackage{graphicx}
\usepackage{amssymb}
\usepackage{epstopdf}
\usepackage{textpos} 
\DeclareMathOperator{\Proj}{Proj}
\DeclareMathOperator{\Tr}{Tr}
\title{Entangled Histories}
\author{Jordan Cotler$^{1,2}$ and Frank Wilczek$^{1,3}$\\
\small\it 1. Center for Theoretical Physics, MIT, Cambridge MA 02139 USA \\
\small\it 2. Stanford Institute for Theoretical Physics, Stanford University, Stanford CA 94305 USA \\
\small \it 3. Origins Project, Arizona State University, Tempe AZ 25287 USA}

\begin{document}
\maketitle


\begin{abstract}
We introduce quantum history states and their mathematical framework, thereby reinterpreting and extending the consistent histories approach to quantum theory.  Through thought experiments, we demonstrate that our formalism allows us to analyze a quantum version of history in which we reconstruct the past by observations.  In particular, we can pass from measurements to inferences about ``what happened'' in a way that is sensible and free of paradox.  Our framework allows for a richer understanding of the temporal structure of quantum theory, and we construct history states that embody peculiar, non-classical correlations in time.
\end{abstract}

Many quantities of physical interest are more naturally expressed in terms of histories than in terms of ``observables'' in the traditional sense, i.e. operators in Hilbert space that act at a particular time.   The accumulated phase $\exp i \int\limits_1^2 \,  dt \, \vec v \cdot \vec A$ of a particle moving in an electromagnetic potential, or its accumulated proper time, are simple examples.   We may ask: Having performed a measurement of this more general, history-dependent sort of observable, what have we learned?   For conventional observables, the answer is that we learn our system is in a particular subspace of Hilbert space, that is the eigenspace corresponding to the observable's measured value.   Here we propose a general framework for formulating and interpreting history-dependent observables.  

Over the last thirty years, the quantum theory of histories has been approached from several directions.  In the 1980's, Griffiths developed a mathematically precise formulation of the Copenhagen interpretation \cite{Griffiths1}.  Griffiths was able to elucidate seemingly paradoxical experiments by enforcing a consistent interpretation of quantum evolution.   Omn\`{e}s, Gell-Mann, Hartle, Isham, and Linden, among others, enriched the mathematics and physics of Griffith's theory of ``consistent histories" \cite{Griffiths2}-\cite{IL3}.  In particular, Gell-Mann and Hartle focused on applying consistent histories to decoherence and quantum cosmology, while Isham and Linden's work has uncovered deep mathematical structure at the foundations of quantum mechanics \cite{Isham1}.

Histories are of course an explicit element of Feynman's path integral.  In the early 1990's, Farhi and Gutmann developed a generalized theory of the path integral, which clarifies the meaning of path integral trajectories for a spin system, or any other system with non-classical features \cite{Farhi-Gutmann}.

Aharonov et. al. developed a formalism for treating systems with multiple sequential pre- and post-selections, called ``multiple-time states" \cite{Aharonov1, Aharonov2}.   This formalism incorporates the interesting possibility of superposing different post-selection schemes, but it does not expose the temporal structure of unitary systems in as full detail as we obtain by splicing time evolution into consistent histories.   

Here we will construct a formal structure that builds on these lines of work, and illuminates the question posed in our first paragraph.  Specifically, we will show how measurements in the present allow us to reconstruct the past evolution of a quantum system.  The possibility arises that there are multiple evolutions of a system which give rise to the same outcome of a measurement in the present, and we analyze how our best description of the past is a quantum superposition of these various evolutions.  Our formalism elucidates that measurement in the present can often force our description of the past to be an \textit{entangled} superposition of evolutions.  Entangled histories appear to capture, in mathematical form, the heuristic concepts of ``parallel universes'' or ``many worlds'' that appear in many discussions of quantum theory: particular time slices will contain orthogonal states (in the conventional sense), which however come together within histories, and at that level can interfere.  Throughout the paper, we analyze several instructive examples, where we apply the formalism to analyze quasi-realistic thought experiments.

\newpage
\section{Mathematics of History States}

\subsection{History Space}

We will work with a vector space that allows theoretical access to the evolution of a system at multiple times.  This vector space is called the history Hilbert space $\mathcal{H}$, and is defined by the tensor product from right to left of the admissible Hilbert spaces of our system at sequential times.  Explicitly, for $n$ times $t_1< \cdots <t_n$, we have
\begin{equation}
\mathcal{H} := \mathcal{H}_{t_n} \odot \cdots \odot \mathcal{H}_{t_1}
\end{equation}
where $\mathcal{H}_{t_i}$ is the admissible Hilbert space at time $t_i$.  Restricting the admissible Hilbert spaces $\mathcal{H}_{t_1}$ and $\mathcal{H}_{t_n}$ corresponds to pre- and post-selection respectively.

In this paper, we will be primarily concerned with history Hilbert spaces defined over a discrete set of times.  It is possible to work with history Hilbert spaces over a continuum of times, but doing so requires the full apparatus of the Farhi-Gutmann path integral \cite{Farhi-Gutmann}.  The history Hilbert space $\mathcal{H}$ is also equipped with  bridging operators $T(t_j, t_i)$, where $T(t_j, t_i) : \mathcal{H}_{t_i} \to \mathcal{H}_{t_j}$.  These bridging operators encode unitary time evolution.

\subsection{History States}

We would like to define a mathematical object that encodes the evolution of our system through time -- a notion of quantum state for history  space -- in a way that supports an inner product and probability interpretation.  One might at first think that an element of $\mathcal{H}$, such as $|\psi(t_n)\rangle \odot \cdots \odot |\psi(t_1)\rangle$ would be the appropriate mathematical object, but a different concept, which removes awkward phases, appears more fruitful.  For us, history states are elements of the linear space  $\Proj(\mathcal{H})$ spanned by projectors from $\mathcal{H} \to \mathcal{H}$.  Henceforth, we will call $\Proj(\mathcal{H})$ the history state space.  

For example, if $\mathcal{H}$ is the history space of a spin-1/2 particle at three times $t_1 < t_2 < t_3$, then an example history state (in history {\it state\/} space) is 
\begin{equation}\label{simplestState}
[z^-] \odot [x^+] \odot [z^+] 
\end{equation}
where we use the notation $[z^+] := |z^+\rangle \langle z^+|$. The history state in Eqn.\,(\ref{simplestState}) can be considered as a quantum trajectory: the particle is spin up in the $z$-direction at time $t_1$, spin-up in the $x$-direction at time $t_2$, and spin-down in the $z$-direction at time $t_3$.  Since $\Proj(\mathcal{H})$ is a complex vector space, another example of a history state is
\begin{equation}\label{simpleSuperposition}
\alpha \, [z^-] \odot [x^+] \odot [z^+]  + \beta \, [z^+] \odot [x^-] \odot [z^+] 
\end{equation}
for complex coefficients $\alpha$ and $\beta$.  The history state in Eqn.\,(\ref{simpleSuperposition}) is a superposition of the history states $[z^-] \odot [x^+] \odot [z^+]$ and $[z^+] \odot [x^-] \odot [z^+]$.  It can be interpreted as meaning that the particle takes \textit{both} quantum trajectories, but with different amplitudes.  

Precise physical interpretation of history states like the one in Eqn.\,(\ref{simpleSuperposition}) requires more structure.  For example, it is not obviously true (and usually is false) that one can measure the particle to take the trajectory $[z^-] \odot [x^+] \odot [z^+]$ with probability proportional to $|\alpha|^2$,  or the other trajectory with probability proportional to $|\beta|^2$.   To discuss probabilities, generalizing the Born rule, we need an inner product.  Furthermore, we have not yet defined which mathematical objects correspond to measurable quantities.

For a history space $\mathcal{H}$ with $n$ times $t_1 < \cdots < t_n$, a general history state takes the form
\begin{equation}
|\Psi) = \sum_{i} \alpha_i [a_i(t_n)] \odot \cdots \odot [a_i(t_1)] 
\end{equation}
where each $[a_i(t_j)]$ is a one-dimensional projector $[a_i(t_j)] : \mathcal{H}_{t_j} \to \mathcal{H}_{t_j}$, and $\alpha_i \in \mathbb{C}$.  We have decorated the history state with a soft ket $|\,\cdot\,)$ which is suggestive of a wave function.  In our theory, a history state is the natural generalization of a wave function, and has similar algebraic properties.  Note that such sums of products of projectors will also accommodate products of hermitian operators more generally.  

\subsection{Inner Product}

In defining a physically appropriate inner product between history states the $K$ operator or ``chain operator" \cite{Griffiths3} defined by
\begin{align}
K|\Psi) &= \sum_{i} \alpha_i K([a_i(t_n)] \odot \cdots \odot [a_i(t_1)]) \\
&= \sum_{i} \alpha_i [a_i(t_n)] T(t_{n},t_{n-1}) [a_i(t_{n-1})] \cdots [a_i(t_2)] T(t_2, t_1) [a_i(t_1)]
\end{align}
where $T(t_j, t_i)$ is the bridging operator associated with the history space, plays a central role.  Note that $K$ maps a history state in $\Proj(\mathcal{H})$ to an operator which takes $\mathcal{H}_{t_1} \to \mathcal{H}_{t_n}$. 

Using the $K$ operator, we equip history states with the positive semi-definite inner product \cite{Griffiths1}
\begin{equation}
(\Phi|\Psi) :=  \Tr \left[ (K|\Phi))^\dagger K|\Psi) \right]
\end{equation}
This inner product induces a semi-norm on history states, and we call $(\Psi|\Psi)$ the {\it weight\/} of $|\Psi)$.  It reflects the probability of $|\Psi)$ occurring.
Note that the inner product is degenerate, in the sense that $(\Psi|\Psi) = 0$ does not imply that $|\Psi) = 0$.  

We say that a history state $|\Psi)$ is normalized if $(\Psi|\Psi)=1$.  If $|\Psi)$ has non-zero weight, then
\begin{equation}
|\overline{\Psi}) = \frac{|\Psi)}{\sqrt{(\Psi|\Psi)}}
\end{equation}
is normalized.  We will use this bar notation throughout the rest of the paper.

At this point an example may be welcome.  If $\mathcal{H}$ is the history space of a spin-1/2 particle at three times $t_1<t_2<t_3$ equipped with a trivial bridging operator $T = \textbf{1}$, then
\begin{equation} \label{KopEx1}
K\left([z^-] \odot [x^+] \odot [z^+] \right) = \frac{1}{2} \, |z^-\rangle \langle z^+| 
\end{equation}
Let us interpret the factor of $1/2$ on the right-hand side of the above equation.  Since the bridging operator for the history space is the identity, any particle which is spin-up in the $z$-direction at time $t_1$ will continue to be in that state at times $t_2$ and $t_3$.  However, the history state in Equation \eqref{KopEx1} is $[z^-] \odot [x^+] \odot [z^+]$, which does not follow the unitary evolution imposed by the bridging operator.  That deviation comes at a cost, which is the amplitude $1/2$.  In general, as we will see, histories which do not follow unitary evolution have a reduced probability of being measured, with the suppression factor proportional to the absolute square of the coefficient generated by the $K$ operator.  In Equation \eqref{KopEx1}, the suppression is a factor of $|1/2|^2 = 1/4$.



\subsection{Families}

We will be interested in subspaces that both admit an orthogonal basis (possibly including history states of zero norm) and contain the complete history state $\textbf{1}_{t_n} \odot \cdots \odot \textbf{1}_{t_1}$ (which corresponds to a system being in a superposition of all possible states at each time).  The orthogonal set of history states which spans such a subspace will be called a {\it family\/} of history states.  More formally:  \\

\noindent \textbf{Definition}   \textit{We say} $\{|\overline{Y}^i)\}$ \textit{is a family of history states if} \\
(1) \,$(\overline{Y}^i | \overline{Y}^j) = 0$ \textit{ for }$i\not = j$ \textit{and} $(\overline{Y}^i | \overline{Y}^i) = 0 \text{ or }1$, \textit{and} \\
(2) \, $\sum_i c_i \, |\overline{Y}^i) = \textbf{1}_{t_n} \odot \cdots \odot \textbf{1}_{t_1}$ \textit{for some complex} $c_i$. \\ \\
\noindent
Requirement (1) is Griffiths'  ``strong consistent histories condition" \cite{Griffiths1}-\cite{Griffiths3}.  

Contrary to earlier work, we see no reason to impose the requirement that history states, regarded as operators, commute.  It is not essential that history states commute since they are not themselves observables.  However, projectors of the form $|\overline{Y}^i)(\overline{Y}^i|$ \textit{are} observables.  By orthogonality, $[|\overline{Y}^i)(\overline{Y}^i|, |\overline{Y}^j)(\overline{Y}^j|] = 0$ for all $i,j$, so commutativity of the corresponding observables is automatic.  We also remark that a family of history states contains at most $\dim(\mathcal{H}_{t_n})\cdot \dim(\mathcal{H}_{t_1})$ history states with non-zero norm \cite{Diosi}.

Now let us work through an example.  Let us consider again the history space of a spin-1/2 particle at three times $t_1<t_2<t_3$ equipped with a trivial bridging operator.  According to our definition,
\begin{align} \label{FirstBasisExample}
|\overline{Y}^1) &= \sqrt{2}\, [z^+]\odot[x^+]\odot[z^+] + \sqrt{2}\, [z^-]\odot[x^-]\odot[z^+] \nonumber \\
|\overline{Y}^2) &= \sqrt{2}\, [z^-]\odot[x^+]\odot[z^+] +\sqrt{2}\, [z^+]\odot[x^-]\odot[z^+] \nonumber \\
|\overline{Y}^3) &= \sqrt{2}\, [z^+]\odot[x^+]\odot[z^-] + \sqrt{2}\, [z^-]\odot[x^-]\odot[z^-] \nonumber \\
|\overline{Y}^4) &= \sqrt{2}\, [z^-]\odot[x^+]\odot[z^-] + \sqrt{2}\, [z^+]\odot[x^-]\odot[z^-]
\end{align}
forms a family of history states.  This family has the curious property that each history state is entangled, in the sense that it is a linear combination of history states which cannot be represented as a product.  Each history state in this family as an entangled quantum trajectory.

Let us explore some of the history states that live in $\text{span}\{|\overline{Y}^1), |\overline{Y}^2), |\overline{Y}^3),|\overline{Y}^4)\}$.  One such history state is $|\overline{\Psi}) = [z^+] \odot [z^+] \odot [z^+]$, which can be written as
\begin{equation} \label{histSup1}
|\overline{\Psi}) = \frac{1}{\sqrt{2}} \, |\overline{Y}^1) + \frac{1}{\sqrt{2}} \, |\overline{Y}^2)
\end{equation}
Equation \eqref{histSup1} implies that a state which at time $t_1$ is spin up in the $z$-direction and evolves in time by the trivial bridging operator can be measured to be the history $|\overline{Y}^1)$ with probability $|1/\sqrt{2}|^2 = 1/2$, or the history $|\overline{Y}^2)$ with probability $|1/\sqrt{2}|^2 = 1/2$.  Later we will outline how to make such a measurement.

Another interesting history state in $\text{span}\{|\overline{Y}^1), |\overline{Y}^2), |\overline{Y}^3),|\overline{Y}^4)\}$ is
\begin{align}
|\Phi) &= \alpha\, [z^+] \odot [z^+] \odot [z^+] + \beta \, [z^-] \odot [z^-] \odot [z^-] \\
&= \frac{\alpha}{\sqrt{2}} \, |\overline{Y}^1) + \frac{\alpha}{\sqrt{2}} \, |\overline{Y}^2) + \frac{\beta}{\sqrt{2}} \, |\overline{Y}^3) + \frac{\beta}{\sqrt{2}} \, |\overline{Y}^4)
\end{align}
which is normalized if $|\alpha|^2 + |\beta|^2 = 1$.  The history state $|\Phi)$ is itself an entangled quantum trajectory.  We will argue that such objects, which might appear exotic, govern concrete measurements.

\subsection{Operators and Observables}

Having provided several examples of history states, we will now briefly discuss operators on histories. We also consider operators $\widehat{A}$ which are linear maps from history states to history states.  In general, any operator of the form
\begin{equation}
\widehat{A} : [\psi(t_n)] \odot \cdots \odot [\psi(t_1)] \longmapsto \sum_i \alpha_i A_i^{t_n} [\psi(t_n)](A_i^{t_n})^\dagger \odot \cdots \odot A_i^{t_1} [\psi(t_1)] (A_i^{t_1})^\dagger
\end{equation}
where all $A_i^{t_j}$ are linear operators, is a linear operator on history state space.  

As in standard quantum theory, not all operators correspond to observables.  History state operators that correspond to observables are those which are both hermitian, and whose eigenvectors can be extended to define a family.  Recall that the latter condition includes the consistent histories condition.  This requirement, which appears necessary for a sensible interpretation of probabilities, excludes many hermitian operators.  

For example, given a family of history states $\{|\overline{Y}^i)\}$, all history state operators of the form
\begin{equation}
\widehat{B} = \sum_i b_i \, |\overline{Y}^i)(\overline{Y}^i |
\end{equation}
for $b_i \in \mathbb{R}$ correspond to observables.  A measurement of the normalized history state $|\overline{\Psi} )$ by the observable $\widehat{B}$ gives the result $b_i$ with probability $| (\overline{\Psi} | \overline{Y}^i )|^2$ (or, for degenerate eigenvalues, the appropriate sum over such terms).  Similar to standard quantum theory, an eigenvalue $b_i$ can be thought of as the readout of a detector which implements the observable $\widehat{B}$ and measures a history state to be $|\overline{Y}^i)$.

For another example, consider the history space of a spin-1/2 particle at two times $t_1<t_2$ equipped with a trivial bridging operator.  We will consider the operator $\sigma_y \odot \sigma_x$ which induces a linear map on history states by
\begin{equation}
\widehat{C} |\Psi) = \sum_i \alpha_i \,\sigma_y [\psi(t_2)] \sigma_y^\dagger \odot  \sigma_x [\psi(t_1)] \sigma_x^\dagger
\end{equation}
The $\widehat{C}$ operator corresponds to measuring the spin-1/2 particle at time $t_1$ in the $x$-basis, and then measuring at time $t_2$ in the $y$-basis.  The eigenhistory states of $\widehat{C}$ form a family, namely $\{\sqrt{2} \, [y^{\pm}]\odot[x^{\pm}]\}$ which are the history state ``outputs" of the sequence of measurements.  More generically, if our history space has a larger number of times, the linear map on history states induced by
\begin{equation}
\textbf{1} \odot \cdots \odot \textbf{1}\odot \sigma_y \odot \textbf{1} \odot \cdots \odot \textbf{1} \odot \sigma_x \odot \textbf{1} \odot \cdots \odot \textbf{1}
\end{equation}
corresponds to measuring at some particular time in the $x$-basis followed by measuring at some later time in the $y$-basis.

We see that measurements at one or more times give rise to families of history states in which the history states are eigenstates of observables (in the traditional sense) at specified times.  

New features appear when we consider more elaborate sets of history observables.  Consider again
a spin-1/2 particle at two times $t_1<t_2$ equipped with a trivial bridging operator.  In addition to $\sigma_y \odot \sigma_x$, we will also consider $\sigma_x \odot \sigma_z$ which induces a linear map on history states by
\begin{align}
\widehat{D} |\Psi) &= \sum_i \alpha_i \,\sigma_x [\psi(t_2)] \sigma_x^\dagger \odot  \sigma_z [\psi(t_1)] \sigma_z^\dagger
\end{align}
The restrictions of $\sigma_y \odot \sigma_x$ and $\sigma_x \odot \sigma_z$ to either $t_1$ or $t_2$ do \textit{not} commute.  However, $\sigma_x \odot \sigma_z$ and $\sigma_z \odot \sigma_y$ themselves \textit{do} commute, which reflects their non-trivial temporal structure.  Thus, $\sigma_x \odot \sigma_z$ and $\sigma_z \odot \sigma_y$ have simultaneous eigenvectors, and correspondingly $\widehat{C}$ and $\widehat{D}$ have simultaneous eigenhistory states which in fact form a family.  The simultaneous eigenvectors of $\sigma_y \odot \sigma_x$ and $\sigma_x \odot \sigma_z$ are
\begin{align}
|\Psi_1\rangle &= - \frac{i}{2} \, |z^+\rangle \odot |z^+\rangle - \frac{1}{2} \, |z^+\rangle \odot |z^-\rangle - \frac{i}{2} \, |z^-\rangle \odot |z^+\rangle + \frac{1}{2} \, |z^-\rangle \odot |z^-\rangle \\
|\Psi_2\rangle &= \frac{i}{2} \, |z^+\rangle \odot |z^+\rangle - \frac{1}{2} \, |z^+\rangle \odot |z^-\rangle + \frac{i}{2} \, |z^-\rangle \odot |z^+\rangle + \frac{1}{2} \, |z^-\rangle \odot |z^-\rangle \\
|\Psi_3\rangle &= - \frac{i}{2} \, |z^+\rangle \odot |z^+\rangle + \frac{1}{2} \, |z^+\rangle \odot |z^-\rangle + \frac{i}{2} \, |z^-\rangle \odot |z^+\rangle + \frac{1}{2} \, |z^-\rangle \odot |z^-\rangle \\
|\Psi_4\rangle &= \frac{i}{2} \, |z^+\rangle \odot |z^+\rangle + \frac{1}{2} \, |z^+\rangle \odot |z^-\rangle - \frac{i}{2} \, |z^-\rangle \odot |z^+\rangle + \frac{1}{2} \, |z^-\rangle \odot |z^-\rangle
\end{align}
and thus the simultaneous eigenhistory states of $\widehat{C}$ and $\widehat{D}$ are
\begin{align}
|\overline{\Psi}_1) &= \sqrt{2}\,|\Psi_1\rangle\langle \Psi_1| \\
|\overline{\Psi}_2) &= \sqrt{2}\,|\Psi_2\rangle\langle \Psi_2| \\
|\overline{\Psi}_3) &= \sqrt{2}\,|\Psi_3\rangle\langle \Psi_3| \\
|\overline{\Psi}_4) &= \sqrt{2}\,|\Psi_4\rangle\langle \Psi_4|
\end{align}
Since these history states are orthonormal and since $\sum_i \frac{1}{\sqrt{2}}\,|\overline{\Psi}_i) = \textbf{1}_{t_2} \odot \textbf{1}_{t_1}$, they constitute a family.

Having developed and exemplified the necessary mathematical machinery, we will now apply our framework to several model systems through quasi-realistic thought experiments.  We will show that the mathematical machinery leads to physically sensible results, and elucidates temporal phenomena in quantum theory which are not transparent in the standard framework of quantum mechanics.

\newpage
\section{Examples and Applications}

\subsection{Mach-Zehnder Interferometer}
Consider the Mach-Zehnder interferometer in Figure 1:
${}$  \\
\begin{center}\label{mach-zehnder}
\includegraphics[scale=0.5]{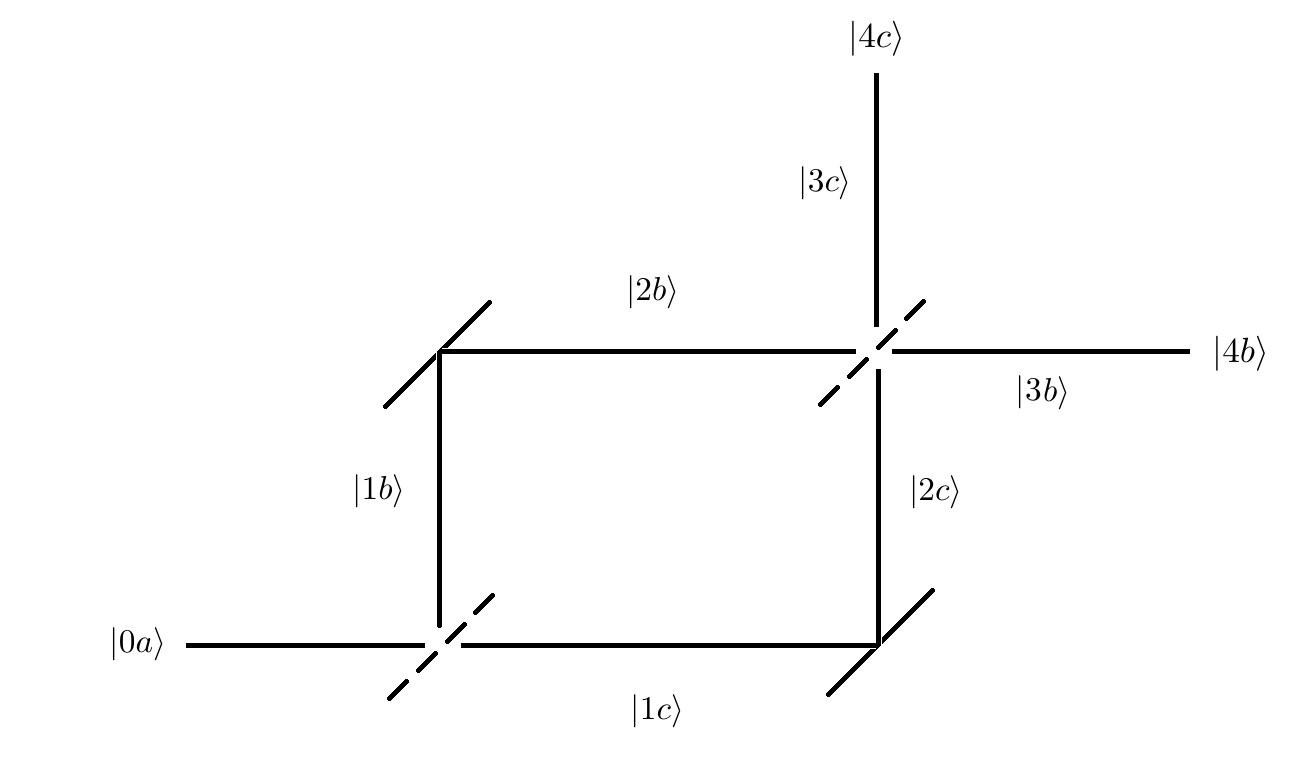}\\
${}$  \\
Figure 1.  Diagram of a Mach-Zehnder interferometer.
\end{center}
${}$ \\
The unitary time evolution of the system is
\begin{align} \label{UnitaryEvol1}
|0a\rangle &\longrightarrow \frac{1}{\sqrt{2}} \left( |1b\rangle + |1c\rangle \right) \longrightarrow \frac{1}{\sqrt{2}} \left( |2b\rangle + |2c\rangle \right)\longrightarrow |3b\rangle \longrightarrow |4b\rangle
\end{align}
which displays interference.  Note that the 50-50 beamsplitters act as the Hadamard matrix $\frac{1}{\sqrt{2}}\begin{bmatrix}
1 & 1 \\
1 & -1 
\end{bmatrix}$ on the spatial modes.  Equation \eqref{UnitaryEvol1} induces a history space of five times $t_0 < t_1 < t_2 < t_3 < t_4$, with a bridging operator which implements the unitary evolution of the system.  Let us work with the family of history states
\begin{align}
|\overline{\alpha}^1) &= 2 \left([4c] \odot \textbf{1}_{t_3} \odot [2b] \odot \textbf{1}_{t_1} \odot [0a] + [4b] \odot \textbf{1}_{t_3} \odot [2c] \odot \textbf{1}_{t_1} \odot [0a]\right) \\
|\overline{\alpha}^2) &= 2 \left([4c] \odot \textbf{1}_{t_3} \odot [2c] \odot \textbf{1}_{t_1} \odot [0a] + [4b] \odot \textbf{1}_{t_3} \odot [2b] \odot \textbf{1}_{t_1} \odot [0a]\right)
\end{align}
where $[0a] = \textbf{1}_{t_0}$ since $\mathcal{H}_{t_0} = \text{span}\{|0a\rangle\}$.  Notice that each history state is an entangled quantum trajectory.  We have
\begin{equation} \label{SumtoUnity1}
\frac{1}{\sqrt{2}}\,|\overline{\alpha}^1) + \frac{1}{\sqrt{2}}\,|\overline{\alpha}^2) = \textbf{1}_{t_4} \odot \textbf{1}_{t_3} \odot \textbf{1}_{t_2} \odot \textbf{1}_{t_1} \odot \textbf{1}_{t_0}
\end{equation}
where the right-hand side corresponds to the history state in which the particle evolves unitarily, because we are not imposing in which state the particle should be at any time.  Equation \eqref{SumtoUnity1} implies that we can measure a particle traveling through the Mach-Zehnder interferometer to be in the history state $|\overline{\alpha}^1)$ with probability $|1/\sqrt{2}|^2 = 1/2$, or $|\overline{\alpha}^2)$ with probability $|1/\sqrt{2}|^2 = 1/2$.  We will now show how, through appropriate coupling to auxiliary qubits, such a measurement can be performed.  

To measure the unitary evolution of the Mach-Zehnder interferometer with respect to the $|\overline{\alpha}^1)$, $|\overline{\alpha}^2)$ family, we couple the system at time $t_0$ to an auxiliary qubit which lives in the space $\text{span}\{|0\rangle, |1\rangle\}$.  We then evolve the original system in time while the auxiliary qubit goes along for the ride, and when appropriate, apply a CNOT gate (treating the auxiliary qubit as the target qubit) so that we can mark histories.  (A similar technique to mark parts of temporal evolution was used by Aharonov et. al. in their work on multiple-time states \cite{Aharonov2}. ) For our purposes, it is essential that we do not tamper with the unitary evolution of the original system by inadvertently imposing additional orthogonality relations  -- i.e., informally, by inadvertently ``collapsing the wavefunction''.  For example, since at time $t_1$ the unitary evolution of the original system gives $\frac{1}{\sqrt{2}} ( |1b\rangle + |1c\rangle)$, it is admissible for the auxiliary qubit to interact with the system according to
\begin{equation}
\frac{1}{\sqrt{2}} \left( |1b\rangle + |1c\rangle \right) \otimes |0\rangle \quad \text{    or     } \quad \frac{1}{\sqrt{2}} \left( |1b\rangle\otimes |0\rangle + |1c\rangle \otimes |1\rangle \right)
\end{equation}
but not according to
\begin{equation}
\frac{1}{2} \, |1b\rangle \otimes |0\rangle + \left(\frac{1}{2}\,|1b\rangle +\frac{1}{\sqrt{2}} \, |1c\rangle \right)\otimes |1\rangle
\end{equation}
since the latter imposes additional orthogonality which decoheres the system.  We emphasize that the system decoheres in the sense that components of the state which which were previously non-orthogonal can no longer interfere with one another due to the presence of the auxiliary qubits which impose extraneous orthogonality.

The desired evolution of the combined Mach-Zehnder-qubit system is as follows:
\begin{align}
|0a\rangle \otimes |0\rangle &\xrightarrow{\,\,\,\,\,T(t_1,t_0) \otimes \textbf{1}\,\,\,\,\,} \frac{1}{\sqrt{2}}\left(|1b\rangle \otimes |0\rangle + |1c\rangle \otimes |0\rangle \right) \\
&\xrightarrow{ T(t_2,t_1) \otimes \textbf{1},\,\, U_1} \frac{1}{\sqrt{2}}\left(|2b\rangle \otimes |0\rangle + |2c\rangle \otimes |1\rangle \right) \\
&\xrightarrow{\,\,\,\,\,T(t_3, t_2)\otimes \textbf{1}\,\,\,\,\,} \frac{1}{\sqrt{2}}\left(\frac{1}{\sqrt{2}}\left(|3b\rangle + |3c\rangle\right) \otimes |0\rangle + \frac{1}{\sqrt{2}}\left(|3b\rangle - |3c\rangle\right) \otimes |1\rangle \right) \\
&\xrightarrow{T(t_4,t_3) \otimes \textbf{1},\,\, U_2} \frac{1}{\sqrt{2}}\left(\frac{1}{\sqrt{2}}\left(|4b\rangle - |4c\rangle\right) \otimes |0\rangle + \frac{1}{\sqrt{2}}\left(|4b\rangle + |4c\rangle\right) \otimes |1\rangle \right) \label{finalState1}
\end{align}
where we have
\begin{align}
U_1 &= |2b\rangle \langle 2b| \otimes |0\rangle \langle 0| + |2b\rangle \langle 2b| \otimes |1\rangle \langle 1| + |2c\rangle \langle 2c| \otimes |0\rangle \langle 1| + |2c\rangle \langle 2c| \otimes |1\rangle \langle 0| \\
U_2 &= |4b\rangle \langle 4b| \otimes |0\rangle \langle 0| + |4b\rangle \langle 4b| \otimes |1\rangle \langle 1| + |4c\rangle \langle 4c| \otimes |0\rangle \langle 1| + |4c\rangle \langle 4c| \otimes |1\rangle \langle 0| 
\end{align}
In this case, if we measure the auxiliary qubit at the final time $t_4$ and detect $|0\rangle$, then the system has been in the history state $|\overline{\alpha}^1)$, whereas if we detect $|1\rangle$, the system has been in the history state $|\overline{\alpha}^2)$.  Note that the probability amplitude of measuring the system to be in either $|\overline{\alpha}^1)$ or $|\overline{\alpha}^2)$ is $1/\sqrt{2}$ which is reflected in Equation \eqref{SumtoUnity1}.  Furthermore, measuring the auxiliary qubit at time $t_4$ collapses the history state of the system to either $|\overline{\alpha}^1)$ or $|\overline{\alpha}^2)$, each with probability $1/2$.

At the final time $t_4$, it is not necessary to measure the auxiliary qubit in the $\{|0\rangle, |1\rangle\}$ basis.  Instead, we could measure the qubit in any other basis, such as the
\begin{equation} \label{MixedBasis1}
\bigg\{|+\rangle = \frac{1}{\sqrt{3}}\,|0\rangle + i \,\sqrt{\frac{2}{3}}\, |1\rangle\,,\quad |-\rangle = \sqrt{\frac{2}{3}}\, |0\rangle - \frac{i}{\sqrt{3}} \,|1\rangle \bigg\}
\end{equation}
basis.  In the $\{|+\rangle, |-\rangle\}$ basis, Equation \eqref{finalState1} takes the form
\begin{align}
&\left[\left(\frac{1}{2\sqrt{3}} - \frac{i}{\sqrt{6}}\right)|4b\rangle - \left(\frac{1}{2\sqrt{3}} + \frac{i}{\sqrt{6}}\right)|4c\rangle\right] \otimes |+\rangle \nonumber \\
&\quad \quad \quad \quad \quad \quad \quad \quad \quad \quad \quad \quad + \left[\left(\frac{1}{\sqrt{6}} + \frac{i}{2\sqrt{3}}\right)|4b\rangle + \left(- \frac{1}{\sqrt{6}} + \frac{i}{2\sqrt{3}}\right)|4c\rangle\right] \otimes |-\rangle
\end{align} 
Measuring $|+\rangle$ at time $t_4$ corresponds to measuring the history state
\begin{equation} \label{basisElmt1}
\frac{1}{\sqrt{3}}\,|\overline{\alpha}^1) + i \, \sqrt{\frac{2}{3}} \,|\overline{\alpha}^2)
\end{equation}
and likewise $|-\rangle$ corresponds to history state
\begin{equation} \label{basisElmt2}
\sqrt{\frac{2}{3}}\,|\overline{\alpha}^1) - \frac{i}{\sqrt{3}}\, |\overline{\alpha}^2)
\end{equation}

Note that Equations \eqref{basisElmt1} and \eqref{basisElmt2} together form a family of history states for the Mach-Zehnder system.  It is  a linear transformation of our original $\{|\overline{\alpha}^1), |\overline{\alpha}^2)\}$ family.  If two families of history states are related by a linear transformation, we say that they are compatible.  As demonstrated in the example above, compatible families have the nice feature that you can measure a history state in one family, or a history state in the other, using the same auxiliary qubits in different bases.  

It is possible to measure a system in differing compatible families sequentially.  For example, say that we want to measure the Mach-Zehnder system with respect to the $\{|\overline{\alpha}^1), |\overline{\alpha}^2)\}$ family and then by the family described by Equations \eqref{basisElmt1} and \eqref{basisElmt2}.  To do this, we tensor two auxiliary qubits to the initial state of the system, and evolve the system and auxiliary qubits in the same manner as before.  At time $t_4$, we end up with
\begin{equation}
\frac{1}{\sqrt{2}}\left(\frac{1}{\sqrt{2}}\left(|4b\rangle - |4c\rangle\right) \otimes |0\rangle \otimes |0\rangle + \frac{1}{\sqrt{2}}\left(|4b\rangle + |4c\rangle\right) \otimes |1\rangle \otimes |1\rangle \right)
\end{equation} 
If we measure the first auxiliary qubit in the $\{|0\rangle, |1\rangle\}$ basis, then the system collapses to either $|\overline{\alpha}^1)$ or $|\overline{\alpha}^2)$ with equal probability.  If we then measure the second qubit in the basis from Equation \eqref{MixedBasis1}, the effect is to measure either the history state $|\overline{\alpha}^1)$ or the history state $|\overline{\alpha}^2)$ in the compatible family described by Equations \eqref{basisElmt1} and \eqref{basisElmt2}.  

\subsection{Some Observations}

The preceding example illustrates that there are aspects of quantum behavior that are not easy to express in the conventional time-evolution picture.   Our theory of history states allows us to manipulate time correlations and time entanglement in a transparent fashion.  

A general CNOT operator takes the form
\begin{equation}
U = \sum_i |i\rangle \langle i| \otimes U_i
\end{equation}
where $\{|i\rangle\}$ is an orthonormal basis for our system of interest at some particular time, and each $U_i$ is a unitary operator that acts on auxiliary qubits.  Such CNOT operators allow us to ``mark" and ``unmark" histories.   By choosing The $U_i$ carefully, we can use this construction to render the entities we have {\it defined\/} as history observables to \textit{be} observable, concretely.   We need only take care to avoid upsetting the unitary evolution of our system of interest by unwittingly imposing orthogonality relations.   

\subsection{Extreme History Entanglement}

In this section we consider an extreme example of history entanglement involving two particles.  We will utilize the history space of two spin-1/2 particles at three times $t_1 < t_2 < t_3$, equipped with a trivial bridging operator.  Then the history states
\begin{align}
|\overline{Z}^1) &= 2\,[z^+, z^+] \odot [x^+, \textbf{1}] \odot [z^+, x^+] + 2\,[z^+, z^-] \odot [x^+, \textbf{1}] \odot [z^-, x^+] \\
|\overline{Z}^2) &= 2\,[z^+, z^+] \odot [x^-, \textbf{1}] \odot [z^+, x^+] + 2\,[z^+, z^-] \odot [x^-, \textbf{1}] \odot [z^-, x^+] \\
|\overline{Z}^3) &= 2\,[z^+, z^+] \odot [x^+, \textbf{1}] \odot [z^+, x^-] + 2\,[z^+, z^-] \odot [x^+, \textbf{1}] \odot [z^-, x^-] \\
|\overline{Z}^4) &= 2\,[z^+, z^+] \odot [x^-, \textbf{1}] \odot [z^+, x^-]  + 2\,[z^+, z^-] \odot [x^-, \textbf{1}] \odot [z^-, x^-] \\
|\overline{Z}^5) &= 2\,[z^+, z^+] \odot [x^+, \textbf{1}] \odot [z^-, x^+] + 2\,[z^+, z^-] \odot [x^+, \textbf{1}] \odot [z^+, x^+] \\
|\overline{Z}^6) &= 2\,[z^+, z^+] \odot [x^-, \textbf{1}] \odot [z^-, x^+] + 2\,[z^+, z^-] \odot [x^-, \textbf{1}] \odot [z^+, x^+] \\
|\overline{Z}^7) &= 2\,[z^+, z^+] \odot [x^+, \textbf{1}] \odot [z^-, x^-] + 2\,[z^+, z^-] \odot [x^+, \textbf{1}] \odot [z^+, x^-] \\
|\overline{Z}^8) &= 2\,[z^+, z^+] \odot [x^-, \textbf{1}] \odot [z^-, x^-] + 2\,[z^+, z^-] \odot [x^-, \textbf{1}] \odot [z^+, x^-] \\
|\overline{Z}^9) &= 2\,[z^-, z^+] \odot [x^+, \textbf{1}] \odot [z^+, x^+] + 2\,[z^-, z^-] \odot [x^+, \textbf{1}] \odot [z^-, x^+] \\
|\overline{Z}^{10}) &= 2\,[z^-, z^+] \odot [x^-, \textbf{1}] \odot [z^+, x^+] + 2\,[z^-, z^-] \odot [x^-, \textbf{1}] \odot [z^-, x^+] \\
|\overline{Z}^{11}) &= 2\, [z^-, z^+] \odot [x^+, \textbf{1}] \odot [z^+, x^-] + 2\,[z^-, z^-] \odot [x^+, \textbf{1}] \odot [z^-, x^-] \\
|\overline{Z}^{12}) &= 2\,[z^-, z^+] \odot [x^-, \textbf{1}] \odot [z^+, x^-] + 2\,[z^-, z^-] \odot [x^-, \textbf{1}] \odot [z^-, x^-] \\
|\overline{Z}^{13}) &= 2\, [z^-, z^+] \odot [x^+, \textbf{1}] \odot [z^-, x^+] + 2\,[z^-, z^-] \odot [x^+, \textbf{1}] \odot [z^+, x^+] \\
|\overline{Z}^{14}) &= 2\,[z^-, z^+] \odot [x^-, \textbf{1}] \odot [z^-, x^+] + 2\,[z^-, z^-] \odot [x^-, \textbf{1}] \odot [z^+, x^+] \\
|\overline{Z}^{15}) &= 2\,[z^-, z^+] \odot [x^+, \textbf{1}] \odot [z^-, x^-] + 2\,[z^-, z^+] \odot [x^+, \textbf{1}] \odot [z^+, x^+] \\
|\overline{Z}^{16}) &= 2\,[z^-, z^+] \odot [x^-, \textbf{1}] \odot [z^-, x^-] + 2\,[z^-, z^-] \odot [x^-, \textbf{1}] \odot [z^+, x^-]
\end{align}
form a family.  

Each history state is an entangled quantum trajectory, and the entanglement encodes novel physical behavior. Consider, specifically,
$$|\overline{Z}^1) = 2\,[z^+, z^+] \odot [x^+, \textbf{1}] \odot [z^+, x^+] + 2\,[z^+, z^-] \odot [x^+, \textbf{1}] \odot [z^-, x^+] $$
This history state exhibits ``time entanglement": Measuring the first particle at time $t_1$ does not determine the state of the second particle until time $t_3$.  We have that the state of particle $1$ at time $t_1$ is the same as the state of particle $2$ at time $t_3$ -- behavior in time similar to that which a Bell state exhibits in space.  Using constructions similar to the ones in the previous section, it is possible to measure such a history state.  Such extreme entanglement exemplifies the possibility of new structures emerging from the quantum theory of history states.


\newpage
\section{Conclusion}

The framework of history states elucidates the temporal structure of quantum theory and makes sense of an expanded class of observables which act on quantum trajectories.  Using our framework, we are able to understand the past as an entangled superposition of time evolutions which are shaped by the outcome of measurements in the present.  To concretely explore temporal entanglement, we have constructed examples of entangled history states with non-classical correlations in time which can be experimentally realized.  It would be very interesting to connect these ideas to the mathematical theory of (classical) inference and causality, which has matured in recent years \cite{Pearl}.

Recently an experiment that demonstrates the existence of entangled histories has been reported \cite{temporalGHZ}.   In this experiment, a (temporally ordered) sequence of measurements was performed, whose result violates an inequality which all non-entangled histories obey.


\subsection*{Acknowledgements}  
Jordan Cotler is supported by the Fannie and John Hertz Foundation and the Stanford Graduate Fellowship program.  Early work was supported by the Undergraduate Research Opportunities Program at the Massachusetts Institute of Technology.  Frank Wilczek's work is supported by the U.S. Department of Energy under grant Contract Number  DE-SC00012567.

\end{document}